\documentclass[11pt]{article}
\usepackage{amssymb}
\usepackage[numbers,sort&compress]{natbib}
\usepackage{amsmath,amsfonts,graphicx,epsfig}
\usepackage{bm}
\usepackage[dvipsnames]{xcolor}
\definecolor{lgray}{gray}{0.35}
\usepackage[colorlinks=true,urlcolor=Gray,linkcolor=lgray,
citecolor=Gray,
             pdfpagelabels=true,hypertexnames=true,
            plainpages=false,naturalnames=false,
             ]{hyperref}
\newcommand{\HCd}{\mathcal{H}}

\def\HCdt0{\tilde{\HCd}_{0}}

\newcommand\rf[1]{(\ref{eq:#1})}
\newcommand\lab[1]{\label{eq:#1}}
\newcommand\nonu{\nonumber}
\newcommand\br{\begin{eqnarray}}
\newcommand\er{\end{eqnarray}}
\newcommand\be{\begin{equation}}
\newcommand\ee{\end{equation}}

\newcommand\foot[1]{\footnotemark\footnotetext{#1}}
\newcommand\lb{\lbrack}
\newcommand\rb{\rbrack}

\newcommand\llb{\left\lbrack}
\newcommand\rrb{\right\rbrack}

\renewcommand\({\left(}
\renewcommand\){\right)}
\newcommand\bgv{\bigg\vert}              %%

\newcommand\bc{\begin{center}}
\newcommand\ec{\end{center}}

\newcommand\partder[2]{\frac{{\partial {#1}}}{{\partial {#2}}}}
\renewcommand\a{\alpha}

\newcommand\eps{\epsilon}
\newcommand\vareps{\varepsilon}

\newcommand\h{\frac{1}{2}}
\renewcommand\k{\kappa}
\renewcommand\l{\lambda}
\renewcommand\L{\Lambda}
\newcommand\m{\mu}
\newcommand\n{\nu}
\newcommand\om{\omega}

\newcommand\vp{\varphi}
\renewcommand\P{\Phi}
\newcommand\pa{\partial}

\newcommand\twomat[4]{\left(\begD{array}{cc}  %%   2x2 matrix  %ESA
{#1} & {#2} \\ {#3} & {#4} \end{array} \right)}

\newcommand\cM{{\mathcal M}}
\newcommand\cN{{\mathcal N}}

\newcommand{\ct}[1]{\cite{#1}}
\newcommand\PRD[3]{{Phys. Rev.} \textbf{D#1}, #3 (#2)}

\newcommand\PRep[3]{{Phys. Reports} \textbf{#1}, #3 (#2)}

\newcommand\IJMPA[3]{{Int. J. Mod. Phys.} \textbf{A#1}, #3 (#2)}
\newcommand\IJMPD[3]{{Int. J. Mod. Phys.} \textbf{D#1}, #3 (#2)}

\newcommand\vpdot{\stackrel{.}{\varphi}}

\newcommand\udot{\stackrel{.}{u}}

\newcommand\adot{\stackrel{.}{a}}

\newcommand\Hdot{\stackrel{.}{H}}
\newcommand\Hddot{\stackrel{..}{H}}

\begin{document}

\setlength\arraycolsep{2pt}

\renewcommand{\theequation}{\arabic{section}.\arabic{equation}}
\setcounter{page}{1}

\begin{titlepage}

\begin{center}

{\huge \bf Dynamically generated inflationary two-field potential via 
non-Riemannian volume forms}

\vskip 2.0cm

{\Large 
D. Benisty$^\text{\scriptsize1,2}$, E. I. Guendelman $^\text{\scriptsize1,2,3}$ 
E. Nissimov $^\text{\scriptsize4}$ and S. Pacheva $^\text{\scriptsize4}$
}

\end{center}

\vskip 0.5cm

\noindent
$^\text{\scriptsize1}${\it Physics Department, Ben-Gurion University of the Negev, 
Beer-Sheva 84105, Israel} \\ 
$^\text{\scriptsize2}${\it Frankfurt Institute for Advanced Studies (FIAS),
Ruth-Moufang-Strasse~1, 60438 Frankfurt am Main, Germany}\\
$^\text{\scriptsize3}${\it Bahamas Advanced Study Institute and Conferences, 
4A Ocean Heights, Hill View Circle, Stella Maris, Long Island, The Bahamas} \\
$^\text{\scriptsize4}${\it Institute for Nuclear Research and Nuclear Energy, 
Bulgarian Academy of Sciences, Sofia, Bulgaria}

\vskip 2.5cm

% \end{center}

\begin{abstract}
We consider a simple model of modified gravity interacting with a single scalar
field $\vp$ with weakly coupled exponential potential within the framework of 
non-Riemannian spacetime volume-form formalism. The specific form of the
action is fixed by the requirement of invariance under global Weyl-scale symmetry.
Upon passing to the physical Einstein frame we show how 
the non-Riemannian volume elements 
create a second canonical scalar field $u$ and dynamically generate a non-trivial
two-scalar-field potential $U_{\rm eff}(u,\vp)$ with two remarkable
features: (i) it possesses a large flat region for 
large $u$ describing a slow-roll inflation; (ii) it has a stable low-lying 
minimum w.r.t. $(u,\vp)$ representing the dark energy density in the 
``late universe''. We study the corresponding two-field slow-roll inflation 
and show that the pertinent slow-roll inflationary curve $\vp = \vp(u)$ 
in the two-field space $(u,\vp)$ has a very small curvature, 
\textsl{i.e.}, $\vp$ changes very little during the inflationary evolution 
of $u$ on the flat region of $U_{\rm eff}(u,\vp)$. Explicit expressions are
found for the slow-roll parameters which differ from those in the
single-field inflationary counterpart. Numerical solutions for the scalar
spectral index and the tensor-to-scalar ratio are derived agreeing
with the observational data.
\end{abstract}

\end{titlepage}

\hypersetup{linktocpage}
%\hrulefill
\noindent\makebox[\linewidth]{\rule{\textwidth}{1.3pt}}
%\vspace{-20pt}
\tableofcontents
\noindent\makebox[\linewidth]{\rule{\textwidth}{1.3pt}}
%\hrulefill

\setcounter{equation}{0}
\section{Introduction}
Developments in cosmology have been inspired by the idea of inflation 
\cite{Starobinsky:1979ty}-\cite{Albrecht:1982wi} which provides an attractive
solution to the the horizon and the flatness problems. It provides in addition 
a framework for sensible calculations of primordial density perturbations 
\cite{Mukhanov:1981xt}-\cite{Guth:1982ec} which becomes more and more 
interesting due to new data from the Cosmic Microwave Background. 
Dynamically generated models of inflation from modified/extended gravity
such as the Starobinsky model \cite{Starobinsky:1980te} 
(see also the reviews \ct{odintsov-1,odintsov-2}) 
still remain viable and produce some of the  best fits to existing observational 
data compared to other inflationary models \cite{Akrami:2018odb}. 

% Two-scalar-field inflation has been found to support background field trajectories 
% that violate the slow-roll and slow-turn conditions. The potential to evade 
% the constraints put forward by the proponents of the string theory swampland
% \cite{Wang:2019tjh}-\cite{Brahma:2019kch}, a proposal to differentiate 
% low energies theories that arise from string theory from those which do not 
% as was done in \cite{Ben-Dayan:2018mhe}. Moreover, the multi-field inflation 
% predeicts different values for the scalar scalar to tensor ratio $r$, 
% that future observations of the CMB polarization will be more constraint 
% \cite{Seljak:1996ti}-\cite{Seljak:1996gy}. However many models has to guess 
% by hand the structure of the potential by hand, and constraint it with the 
% corresponding data. The non-Rimenian volume form formalism represents a 
% mechanism which dynamically generated the scalar potential from global 
% scale invariance symmetry.  

%% 4D-GB2 - Ind-Inflat
Another efficient way to produce consistent inflationary models dynamically 
from modified/extended gravitational theories 
is based on employing alternative non-Riemannian spacetime volume-forms, 
\textsl{i.e.}, metric-independent generally covariant volume elements,
in the coresponding Lagrangian actions instead of 
the canonical Riemannian volume element given by the square-root of the
determinant of the Riemannian metric $\sqrt{-g} \equiv \sqrt{-\det\Vert g_{\m\n}\Vert}$
(for recent applications of this idea, see 
\cite{Benisty:2018fja,Benisty:2019tno}). The method of
metric-independent volume elements was
originally proposed in % \ct{TMT-orig-0,Hehl,TMT-orig-1,TMT-orig-2,5thforce}, 
\ct{TMT-orig-0}-\ct{5thforce}) with a subsequent concise geometric formulation
in \ct{susyssb-1,grav-bags}.

%% 4D-GB2 - Ind-Inflat
The non-Riemannian volume element formalism was used as a basis for constructing 
(i) a series of extended gravity-matter models describing unified dark energy and 
dark matter scenario \ct{dusty,dusty-2};
(ii) quintessential cosmological models with gravity-assisted and inflaton-assisted
dynamical suppression (in the ``early'' universe) or dynaamical generation (in the
post-inflationary universe) of electroweak spontaneous symmetry
breaking and charge confinement \ct{grf-essay,varna-17,bpu-10}; 
(iii) a novel mechanism for the supersymmetric Brout-Englert-Higgs effect in 
supergravity \ct{susyssb-1}.

%%%%%%%
Let us briefly recall the essence of the non-Riemannian volume-form (volume
element) formalism. In integrals over differentiable manifolds (not necessarily 
Riemannian one, so {\em no} metric is needed) volume-forms are given by 
nonsingular maximal rank differential forms $\om$:
\br
\int_{\cM} \om \bigl(\ldots\bigr) = \int_{\cM} dx^D\, \Omega \bigl(\ldots\bigr)
\;,
\nonu \\
\om = \frac{1}{D!}\om_{\m_1 \ldots \m_D} dx^{\m_1}\wedge \ldots \wedge dx^{\m_D}
\quad ,\quad
\om_{\m_1 \ldots \m_D} = - \vareps_{\m_1 \ldots \m_D} \Omega \; ,
% \;\; ,\;\;
% dx^{\m_1}\wedge \ldots \wedge dx^{\m_D} = \vareps^{\m_1 \ldots \m_D}\,  dx^D \; ,
% \lab{omega-3}
\lab{omega-1}   
\er
(our conventions for the alternating symbols $\vareps^{\m_1,\ldots,\m_D}$ and
$\vareps_{\m_1,\ldots,\m_D}$ are: $\vareps^{01\ldots D-1}=1$ and
$\vareps_{01\ldots D-1}=-1$).
The volume element % (integration measure density) 
$\Omega$ transforms as scalar
density under general coordinate reparametrizations.

In Riemannian $D$-dimensional spacetime manifolds a standard generally-covariant 
volume-form is defined through the ``D-bein'' (frame-bundle) canonical one-forms 
$e^A = e^A_\m dx^\m$ ($A=0,\ldots ,D-1$):
\be
\om = e^0 \wedge \ldots \wedge e^{D-1} = \det\Vert e^A_\m \Vert\,
dx^{\m_1}\wedge \ldots \wedge dx^{\m_D} \quad  
% \nonu \\
\longrightarrow \quad
\Omega = \det\Vert e^A_\m \Vert = \sqrt{-\det\Vert g_{\m\n}\Vert} \; .
\lab{omega-riemannian}
\ee

To construct modified gravitational theories as alternatives to ordinary
standard theories in Einstein's general relativity, instead of $\sqrt{-g}$ 
we can employ one or more alternative {\em non-Riemannian} 
volume element(s) as in \rf{omega-1} given by non-singular {\em exact} $D$-forms
$\om = d A$ where:
\be
A = \frac{1}{(D-1)!} A_{\m_1\ldots\m_{D-1}}
dx^{\m_1}\wedge\ldots\wedge dx^{\m_{-1}} \quad \longrightarrow \quad
% \nonu \\
% \lab{B-form}
% \ee
% so that the {\em non-Riemannian} volume element reads:
% \be
% \longrightarrow \quad  
\Omega \equiv \Phi(A) =
\frac{1}{(D-1)!}\vareps^{\m_1\ldots\m_D}\, \pa_{\m_1} A_{\m_2\ldots\m_D} \; .
\lab{Phi-D}
\ee
Thus, a non-Riemannian volume element is defined in terms of
the (scalar density of the) dual field-strength of an auxiliary rank 
$D-1$ tensor gauge field $A_{\m_1\ldots\m_{D-1}}$. 

%%%%%%%%%%%%%%%%%%%%%%%%%%%%%%%%%%%%%%%%%%%%%%%%%%%%%%%%%%%%%%%%%%%%%%%%%%%%%%%%%%%%%%%%%%%%%%%%
%%%%%%%%%%%%%%%%%%%%%%%%%%%%%%%%%%%%%%%%%%%%%%%%%%%%%%%%%%%%%%%%%%%%%%%%%%%%%%%%%%%%%%%%%%%%%%%%
\section{Modified Gravity-Matter Model with Non-Riemannian Volume Elements}

Let us consider the following specific modified model of gravity 
% in the second order (metric) formulation 
coupled to a real scalar field $\vp$, constructed within the non-Riemannian
spacetime volume-form formalism and involving several non-Riemannian volume
elements, with an action (for simplicity we are using units with the Newton constant 
$G_{\rm Newton} = 1/16\pi$):
\be
S = \int d^4 x \Bigl\{ \P_1 (A) \Bigl\lb R + L^{(1)} 
- 2\L_0 \frac{\P_1 (A)}{\sqrt{-g}}\Bigr\rb
% \nonu \\
+ \P_2 (B)\Bigl\lb L^{(2)} + \frac{\P_0 (C)}{\sqrt{-g}}\Bigr\rb\Bigr\} \;.
\lab{NRVF-1}
\ee
Here the following notations are used:

%%%%%%%%%%%%%%%
\begin{itemize}
\item
$R$ is the scalar curvature of the Riemannian spacetime metric $g_{\m\n}$; 
$\sqrt{-g} \equiv \sqrt{-\det\Vert g_{\m\n}\Vert}$;
\item
We employ three metric-independent volume-elements $\P_1 (A),\,\P_2 (B),\,\P_0 (C)$
built in terms of auxiliary rank 3 antisymmetric tensor gauge fields:
\be
\P_1 (A) \equiv \frac{1}{3!}\vareps^{\m\n\k\l} \pa_\m A_{\n\k\l} \quad,\quad
\P_2 (B) \equiv \frac{1}{3!}\vareps^{\m\n\k\l} \pa_\m B_{\n\k\l} \quad, \quad
% \nonu \\
\P_0 (C) \equiv \frac{1}{3!}\vareps^{\m\n\k\l} \pa_\m C_{\n\k\l} \; .
\phantom{aaaaaa}
\lab{Phi-D4}
\ee
\item
The matter Lagrangian of a single scalar $\vp$ is given in two pieces:
\be
L^{(1)} = -\h g^{\m\n} \pa_\m \vp \pa_\n \vp + f_1 e^{-\a\vp} \quad ,\quad
L^{(2)} = - f_2 e^{-2\a\vp} \; ,
\lab{L-1-2}
\ee
where $f_{1,2}$ and $\a$ are dimensionful (positive) coupling parameters;
$\a$ {\em will be considered small parameter}.
\item
$\L_0$ is dimensionfull parameter to be identified as energy density scale
of the inflationary universe' epoch.
\end{itemize}
%%%%%%%%%%%%%%%

The specific form of the action \rf{NRVF-1} is dictated by the requirement
about global Weyl-scale invariance\foot{Scale invariance played an important role 
since the original papers on the non-canonical volume-form formalism 
% where also fermions were included 
\ct{TMT-orig-1}. Also let us note that spontaneously broken dilatation symmetry
models constructed along these lines are free of the Fifth Force Problem
\ct{5thforce}.} under:
\be
g_{\m\n} \to \l g_{\m\n} \quad,\quad \vp \to \vp + 1/\a\,\log\l \quad,\quad
% \phantom{aaaaaaaaaa}% \G^\m_{\n\l} \to \G^\m_{\n\l} 
% \nonu \\
A_{\m\n\k} \to \l A_{\m\n\k} \quad ,\quad B_{\m\n\k} \to \l^2 B_{\m\n\k} 
\quad ,\quad C_{\m\n\k} \to C_{\m\n\k} \; .
\lab{scale-transf}
\ee

The modified gravitational theories of the type \rf{NRVF-1}, when formulated
within the first-order (Palatini) formalism, possess the following characteristic 
feature. The auxiliary rank 3 tensor gauge fields defining all non-Riemannian 
volume-elements \rf{Phi-D4} are almost {\em pure-gauge} degrees of freedom, 
\textsl{i.e.} 
they {\em do not} introduce any additional propagating gravitational degrees of 
freedom when passing to the physical Einstein frame except for few discrete degrees 
of freedom with conserved canonical momenta appearing as arbitrary integration 
constants. This has been explicitly shown within the canonical Hamiltonian treatment 
(see Appendices A in Refs.\ct{grav-bags,grf-essay}). 

Unlike Palatini formalism, when we treat \rf{NRVF-1} in the second order
(metric) formalism, while passing to the physical Einstein frame via
conformal transformation:
\be
g_{\m\n} \to {\bar g}_{\m\n}= \chi_1 g_{\m\n} \quad ,\quad
\chi_1 \equiv \frac{\P_1 (A)}{\sqrt{-g}} \; ,
\lab{g-bar}
\ee
the first non-Riemannian volume element $\P_1 (A)$ in \rf{NRVF-1} is not any more 
(almost) ``pure gauge'', but creates a new dynamical canonical scalar field 
$u$ via $\chi_1 = \exp{\frac{u}{\sqrt{3}}}$. In Ref.\ct{Benisty:2019tno} we
have shown how, starting from a purely gravitational model within the
non-Riemannian volume element formalism without any coupling to matter
fields (\textsl{i.e.}, starting with an action \rf{NRVF-1} where $L^{(1,2)} = 0$ 
identically), the passage to the physical Einstein frame produces a
non-trivial inflationary potential for the dynamically created scalar field $u$. 
Here below, this idea is extended to produce dynamically a non-trivial 
two-field inflationary potential starting from the single-scalar-field action 
\rf{NRVF-1}.

The equations of motion upon variation of \rf{NRVF-1} w.r.t. the auxiliary 
tensor gauge fields $A_{\m\n\l},\, B_{\m\n\l}\,, C_{\m\n\l}$ yield, respectively, 
%%%%%%%%%%  CHANGE-BEGIN  %%%%%%%%%%%%
the following solutions:
\br
R  -\h g^{\m\n} \pa_\m \vp \pa_\n \vp + f_1 e^{-\a\vp}
- 4\L_0 \frac{\P_1 (A)}{\sqrt{-g}} = - M_1 \equiv {\rm const} \; ,
\lab{A-eq} \\
- f_2 e^{-2\a\vp} + \frac{\P_0 (C)}{\sqrt{-g}} = - M_2 \equiv {\rm const} \;\; , \;\; 
\frac{\P_2 (B)}{\sqrt{-g}} = \chi_2 \equiv {\rm const} \; .
\lab{B-C-eq}
\er
Let us note that Eqs.\rf{B-C-eq} are (non-dynamical) constraints, whereas
Eq.\rf{A-eq} is a dynamical relation -- it contains second-order time
derivatives inside $R$.

In \rf{A-eq}-\rf{B-C-eq}  $M_1 , M_2$ and  $\chi_2$ are (dimensionful and 
dimensionless, respectively) free integration constants. 
The appearance of $M_1 , M_2$ indicate
spontaneous breaking of global Weyl symmetry \rf{scale-transf}.

Accordingly, the equations of motion w.r.t. $g_{\m\n}$ taking into account
Eqs.\rf{B-C-eq} read:
\be
R_{\m\n} - \L_0 \chi_1 g_{\m\n}
+ \frac{1}{\chi_1}\bigl( g_{\m\n} \Box{\chi_1} - \nabla_\m \nabla_\n \chi_1\Bigr)
- \frac{\chi_2}{\chi_1}\( f_2 e^{-2\a\vp} + M_2\) g_{\m\n} = 0 \; ,
\lab{einstein-like}
\ee
with $\chi_1$ as in \rf{g-bar}.
% Performing conformal transformation \rf{g-bar} on the r.h.s. of
% \rf{einstein-like} using the conformal transformation rule for $R_{\m\n}$
%(see \textsl{e.g.} Ref.\ct{dabrowski}; :

We now transform Eqs.\rf{einstein-like} via the
conformal transformation \rf{g-bar} and show that the transformed equations
acquire the standard form of Einstein equations w.r.t. the 
new ``Einstein-frame'' metric ${\bar g}_{\m\n}$. To this end we will use 
the known formulas for the conformal transformations of $R_{\m\n}$ and
$\Box\Psi$ for some arbitrary scalar field, in particular 
$\Psi \equiv \chi_1$ (see \textsl{e.g.} Ref.\ct{dabrowski}; bars indicate 
magnitudes in the ${\bar g}_{\m\n}$-frame):
\be
R_{\m\n}(g) = R_{\m\n}(\bar{g}) - 3 \frac{{\bar g}_{\m\n}}{\chi_1}
{\bar g}^{\k\l} \pa_\k \chi_1^{1/2} \pa_\l \chi_1^{1/2} 
+ \chi_1^{-1/2}\bigl({\bar \nabla}_\m {\bar \nabla}_\n \chi_1^{1/2} +
{\bar g}_{\m\n} {\bar{\Box}}\chi_1^{1/2}\bigr) \; ,
\lab{dabrowski-1} 
\ee
and
\be
\Box \chi_1 = \chi_1 \Bigl({\bar{\Box}}\chi_1 
- 2{\bar g}^{\m\n} \frac{\pa_\m \chi_1^{1/2} \pa_\n \chi_1}{\chi_1^{1/2}}\Bigr)
\; .
\lab{dabrowski-2}    
\ee
Indeed, using \rf{dabrowski-1}-\rf{dabrowski-2} and taking into account all
relations \rf{A-eq} and \rf{B-C-eq}, then
Eqs.\rf{einstein-like} are rewritten in the standard Einstein's form 
of w.r.t. the ``Einstein-frame'' metric ${\bar g}_{\m\n}$:
%%%%%%%%%%  CHANGE-END  %%%%%%%%%%%%
\be
R_{\m\n}(\bar{g}) - \h {\bar g}_{\m\n} R(\bar{g}) =
% \nonu \\
\h \Bigl\lb \pa_\m \vp \pa_\n \vp + \pa_\m u \pa_\n u
% \nonu \\
- {\bar g}_{\m\n}\bigl(\h {\bar g}^{\k\l} \pa_\k \vp \pa_\l \vp
+ \h {\bar g}^{\k\l} \pa_\k u \pa_\l u + U_{\rm eff} (u,\vp)\bigr)\Bigr\rb \; ,
\lab{EF-eqs}
\ee
where we have redefined:
\be
\P_1 (A)/\sqrt{-g}\equiv \chi_1 = \exp{\bigl\{u/\sqrt{3}\bigr\}}
\lab{u-def}
\ee
in order to obtain a canonically normalized kinetic term for the scalar
field $u$, and where:
\br
U_{\rm eff} (u,\vp) = 2 \L_0 - M_1 \exp{\bigl\{-\frac{u}{\sqrt{3}}\bigr\}} 
+ \chi_2 M_2 \exp{\bigl\{-2 \frac{u}{\sqrt{3}}\bigr\}} 
\nonu \\
- f_1 \exp{\bigl\{-\frac{u}{\sqrt{3}}-\a\vp\bigr\}}
+ \chi_2 f_2 \exp{\bigl\{-2 \bigl(\frac{u}{\sqrt{3}}+\a\vp\bigr)\bigr\}} \; .  
\lab{U-eff}
\er

%%%%%%%%%%%%%%%%%%%%%%%%
\begin{figure*}[t!]
\centering
\includegraphics[width=0.7\textwidth]{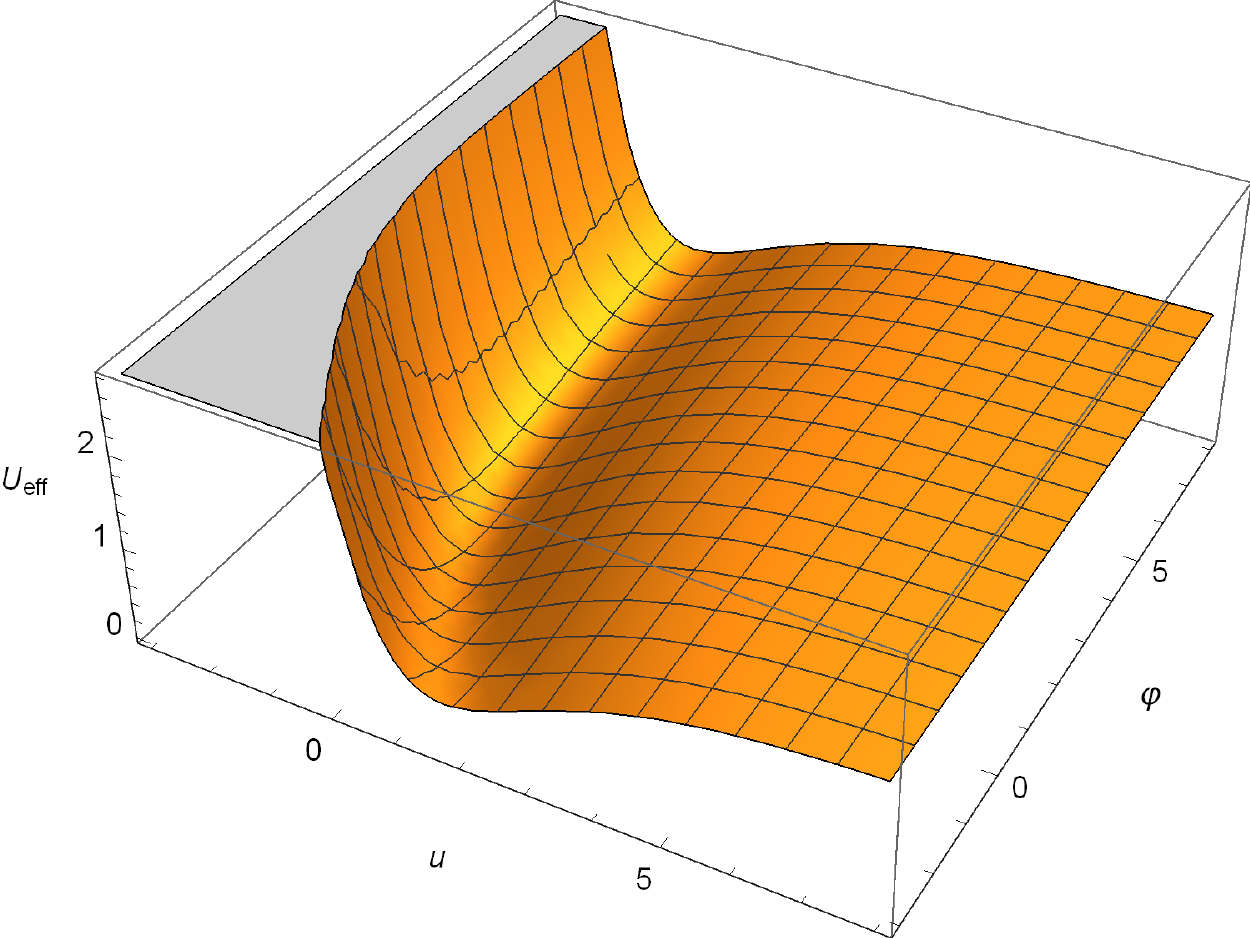}
\includegraphics[width=0.7\textwidth]{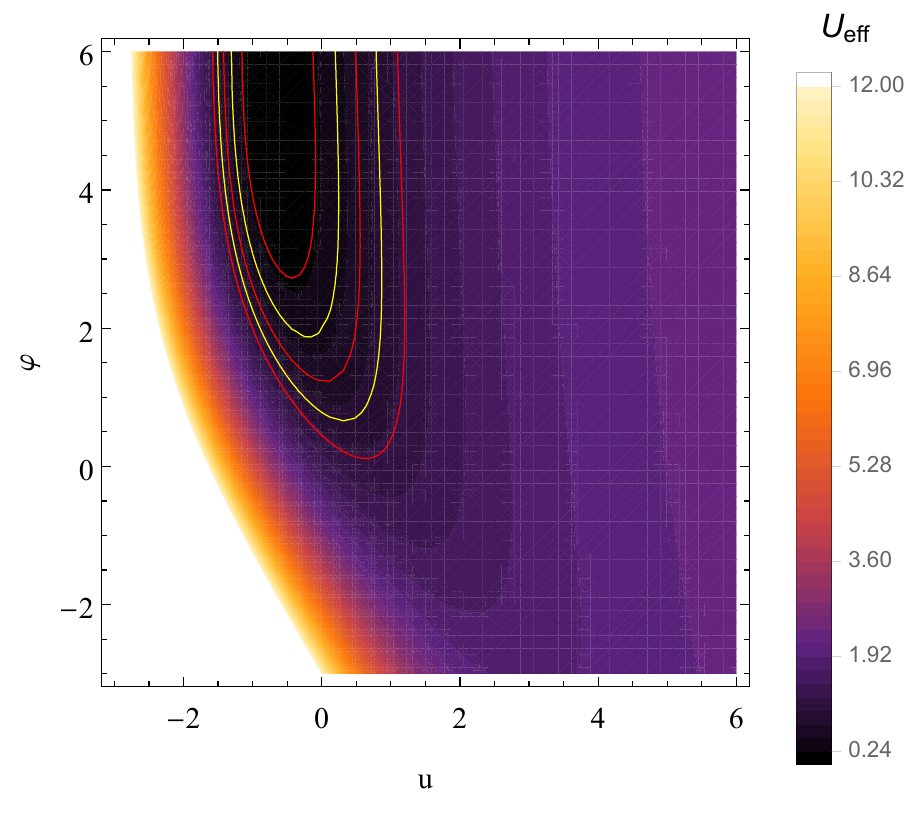}
\caption{The effective potential $U_{\rm eff}(u,\vp)$ \rf{U-eff}
in the physical Einstein frame. The upper panel represents the two-dimensional
graph of the potential,  % of the function $U_{\rm eff}(u,\vp)$,
whereas the lower panel depicts topographically the surface 
$U_{\rm eff} = U_{\rm eff}(u,\vp)$ from above. 
The point of the stable minimum of the potential can be seen in the right panel.}
\label{fig1}
\end{figure*}
%%%%%%%%%%%%%%%%%%%%%%%

The corresponding Einstein-frame action reads:
\be
S_{\rm EF} = \int d^4 x \sqrt{-{\bar g}} \Bigl\lb R({\bar g}) 
% - \h {\bar g}^{\m\n}\pa_\m \vp \pa_\n \vp 
% \nonu\\
- \h {\bar g}^{\m\n} \pa_\m \vp \pa_\n \vp - \h {\bar g}^{\m\n}\pa_\m u \pa_\n u  
- U_{\rm eff} (u, \vp) \Bigr\rb\; .
\lab{EF-action}
\ee
%%%%%%%%%%  CHANGE-BEGIN  %%%%%%%%%%%%
Let us note that we could obtain the Einstein-frame action \rf{EF-action}
directly from the initial action \rf{NRVF-1} in the non-Riemannian volume-form
formalism upon performing there the conformal transformation from $g_{\m\n}$
to ${\bar g}_{\m\n}$ \rf{g-bar}. To this end one has to use the known
formula for the conformal transformation \rf{g-bar} of the scalar curvature 
$R \equiv R(g)$ (\textsl{e.g.}, Ref.\ct{dabrowski}):
\be
R(g) = \chi_1 \Bigl\lb R({\bar g}) + 6 \frac{{\bar{\Box}}\chi_1^{1/2}}{\chi_1^{1/2}}
-12 {\bar g}^{\m\n} \frac{\pa_\m \chi_1^{1/2} \pa_\n \chi_1^{1/2}}{\chi_1}\Bigr\rb 
\; ,
\lab{dabrowski-3}
\ee
and then repeat the same steps as in Ref.\ct{Benisty:2018fja}, 
where such a procedure has
been applied to a simplified version of the original action \rf{NRVF-1}.
%%%%%%%%%%  CHANGE-END  %%%%%%%%%%%%

Let us particularly stress that the Einsten-frame action \rf{EF-action}
contains a dynamically created canonical scalar field $u$ entering a 
non-trivial effective two-field scalar potential $U_{\rm eff}(u,\vp)$ \rf{U-eff} 
-- both are dynamically produced 
by the initial non-Riemannian volume elements in \rf{NRVF-1} due to the
appearance of the free integration constants $M_1, M_2, \chi_2$ 
in their respective equations of motion \rf{A-eq}-\rf{B-C-eq}.

The form of $U_{\rm eff}(u,\vp)$ is graphically depicted on Fig.1 .

%%%%%%%%%%%%%%%%%%%%%%%%%%%%%%%%%%%%%%%%%%%%%%%%%%%%%%%%%%%%%%%%%%%%%%%%%%%%%%%%%%%%%%%%%%%%%%%%
%%%%%%%%%%%%%%%%%%%%%%%%%%%%%%%%%%%%%%%%%%%%%%%%%%%%%%%%%%%%%%%%%%%%%%%%%%%%%%%%%%%%%%%%%%%%%%%%

\section{Applications to Cosmology}

From cosmological point of view the dynamically generated two-field scalar potential 
$U_{\rm eff}(u,\vp)$ \rf{U-eff} possesses the following relevant features:

\begin{itemize}
\item
For large positive $u$, $U_{\rm eff}(u,\vp) \simeq 2\L_0$, \textsl{i.e.},
almost flat region regardless of $\vp$. 
\item
$U_{\rm eff}(u,\vp)$ has a stable minimum at
certain finite values $(u=u_{\rm min}, \vp = \vp_{\rm min})$.
\end{itemize}

Indeed, calculating the extremal point(s): 
% $\partder{U_{\rm eff}}{u}=0$ and $\partder{U_{\rm eff}}{\vp}=0$ we find:
\be
\partder{U_{\rm eff}}{u}=0 \;\;, \;\;\partder{U_{\rm eff}}{\vp}=0 \;\;\; \to \;\;\;
\exp\{-u_{\rm min}/\sqrt{3}\}=\frac{M_1}{2\chi_2 M_2} \;\; ,\;\;
\exp\{-\a\vp_{\rm min}\} = \frac{f_1 M_2}{f_2 M_1} \; ,
\lab{crit-point}
\ee
we find for the matrix of second derivatives of $U_{\rm eff}(u,\vp)$
at the extremal point $(u_{\rm min},\vp_{\rm min})$:
\be
\frac{\pa^2 U_{\rm eff}}{\pa u^2} \bgv_{\rm min} = 
\frac{1}{3}\Bigl\lb \frac{M_1^2}{2\chi_2 M_2} + \frac{f_1^2}{2\chi_2 f_2}\bigr\rb
\quad ,\quad 
\frac{\pa^2 U_{\rm eff}}{\pa \vp^2} \bgv_{\rm min} =
\sqrt{3}\a \;\frac{\pa^2 U_{\rm eff}}{\pa u \pa\vp}\bgv_{\rm min} = 
\a^2 \frac{f_1^2}{2\chi_2 f_2} \; ,
\lab{hessian}
\ee
which is manifestly positive definite (both the diagonal elements as well as 
its determinant  
$\frac{\a^2}{3} \bigl(f_1^2/\chi_2 f_2\bigr)\,\bigl(M_1^2/\chi_2 M_2\bigr)$
are all positive).

The flat region of $U_{\rm eff} (u,\vp)$ for large positive $u$ corresponds to
``early'' universe' slow-roll inflationary evolution with energy scale $2\L_0$. On the
other hand, the region around the stable minimum at $(u_{\rm min},\vp_{\rm min})$ 
\rf{crit-point} corresponds to ``late'' universe' evolution where the minimum 
value of the potential:
\be
U_{\rm eff} (u_{\rm min},\vp_{\rm min})= 2\L_0 - \frac{M_1^2}{4\chi_2 M_2} 
- \frac{f_1^2}{4\chi_2 f_2} \equiv 2 \L_{\rm DE}
\lab{DE-value}
\ee
is the dark energy density value \cite{Angus:2018tko,Zhang:2018gbq,Aghamousa:2016zmz}. 
Thus, to conform to the observational data $\L_0 \sim 10^{-8} M_{Pl}^4$ and
$\L_{\rm DE} \sim 10^{-122} M_{Pl}^4$ we can choose for instance::
\be
M_{1,2} \sim f_{1,2} \sim 10^{-8} M_{Pl}^4
\quad,\quad \chi_2 \sim 1 \quad , \quad {\rm or} \quad
M_{1,2} \sim 10^{-8} M_{Pl}^4 \;\; ,\;\; f_{1,2} \sim M_{EW}^4 \;.
\lab{orders}
\ee
where $M_{EW}\sim 10^{-16} M_{Pl}$ is the electroweak scale and
$M_{Pl}$ is the Planck mass.

Let us now consider reduction of the Einstein-frame action \rf{EF-action} to the
Friedmann-Lemaitre-Robertson-Walker (FLRW) setting with metric
% $ds^2 = - \cN^2 dt^2 + a(t)^2 d{\vec x}^2$, and with $u=u(t)$, $\vp=\vp(t)$.
$ds^2 = - dt^2 + a(t)^2 d{\vec x}^2$, and with $u=u(t)$, $\vp=\vp(t)$.

On the flat region of $U_{\rm eff}(u,\vp) \equiv V(u.\vp)$ for large
positive $u$ (henceforth we
will use the notation $V(u,\vp)$ for simplicity) the slow-roll dynamics:
\br
% 6 H^2 \simeq V(u,\vp) \quad, \quad 3H\vpdot +\partder{V}{\vp} \simeq 0 \quad, \quad
% 3H\udot +\partder{V}{u} \simeq 0 \; ,
6 H^2 \simeq V(u,\vp) \quad, \quad 3H\vpdot + V_{\vp} \simeq 0 \quad, \quad
3H\udot + V_u \simeq 0 \; ,
\lab{slow-roll} \\
V_{\vp} \equiv \partder{V}{\vp} = \a \llb f_1 e^{-u/\sqrt{3} -\a\vp} 
-2 \chi_2 f_2 e^{-2(u/\sqrt{3} +\a\vp)}\rrb \; ,
\lab{V-vp} \\
V_u \equiv \partder{V}{u} = \frac{1}{\sqrt{3}} \llb M_1 e^{-u/\sqrt{3}}
- 2 \chi_2 M_2 e^{-2u/\sqrt{3}} + f_1 e^{-u/\sqrt{3} -\a\vp} 
-2 \chi_2 f_2 e^{-2(u/\sqrt{3} +\a\vp)}\rrb \; ,
\lab{V-u}
\er
where $H$ denotes the Hubble parameter $H=\adot/a$, 
defines an inflationary trajectory in the two-field space $u=u(t),\,\vp=\vp(t)$ 
which can be described as a curve $\vp = \vp (u)$ satisfying:
\be
\frac{d\vp}{du} = \frac{V_\vp}{V_u} = 
\sqrt{3} \a \,\frac{f_1 e^{-u/\sqrt{3} -\a\vp(u)} 
-2 \chi_2 f_2 e^{-2(u/\sqrt{3} +\a\vp(u))}}{M_1 e^{-u/\sqrt{3}}
- 2 \chi_2 M_2 e^{-2u/\sqrt{3}} + f_1 e^{-u/\sqrt{3} -\a\vp(u)} 
-2 \chi_2 f_2 e^{-2(u/\sqrt{3} +\a\vp(u))}} \; .
\lab{vp-u-curve}
\ee

Assuming $\a$ small, Eq.\rf{vp-u-curve} can be solved as:
\be
\vp(u) = \vp_0 + \a \vp_1 (u) + {\rm O}(\a^2) \; ,
\lab{vp-sol}
\ee
where:
\br
\vp_1(u) = 3 y_0 \Bigl\lb \(\frac{f_1}{M_1+f_1 y_0} - 
\frac{f_2 y_0}{M_2+f_2 y_0^2}\) \log\Bigl( M_1+f_1 y_0 -
2\chi_2 (M_2+f_2 y_0^2)e^{-u/\sqrt{3}}\Bigr) 
\nonu \\
+ \; \frac{f_1}{\sqrt{3}(M_1+f_1 y_0)}\, u\Bigr\rb  \; , \phantom{aaaaaaaa}
\lab{vp-sol-1}
\er
with $y_0 \equiv e^{-\a\vp_0}$ and $\vp_0$ being an integration constant.
Comparing \rf{vp-sol} with the explicit expression \rf{vp-sol-1} requires
for consistency in the inflationary region (flat region for large positive
$u$) that the integration constant $\vp_0$ must also be large positive.

We notice that $V \equiv U_{\rm eff}(u,\vp)$ \rf{U-eff} and $V_\vp$ \rf{V-vp},
$V_u$ \rf{V-u} all depend on $(u,\vp)$ only through 
$\bigl(x \equiv e^{-u/\sqrt{3}}, y \equiv e^{-\a\vp}\bigr)$, therefore it
will be convenient to consider them as functions of $(x,y)$:
\br
U_{\rm eff}(u,\vp) \equiv V(x,y)= 2\L_0 - M_1 x + {\bar M}_2 x^2 
- f_1 xy + {\bar f}_2 x^2 y^2 \; ,
\nonu \\
V_\vp (x,y) = \a \bigl\lb f_1 xy - 2{\bar f}_2 x^2 y^2 \bigr\rb \;\; ,\;\;
V_u(x,y) = \frac{1}{\sqrt{3}}\bigl\lb M_1 x - 2{\bar M}_2 x^2 
+ f_1 xy - 2{\bar f}_2 x^2 y^2 \bigr\rb \; ,
% \quad ,\quad 
% {\bar M}_2 \equiv \chi_2 M_2 \; ,\; {\bar f}_2 \equiv \chi_2 f_2 \; .
\lab{V-xy-2}
\er
(here ${\bar M}_2 \equiv \chi_2 M_2 \; ,\; {\bar f}_2 \equiv \chi_2 f_2$).
Accordingly, the inflationary curve $\vp = \vp(u)$ will be represented as a
curve $y=y(x)$ in $(x,y)$ space satisfying the counterpart of Eq.\rf{vp-u-curve}:
\be
\frac{dy}{dx}=\sqrt{3}\a \,\frac{y}{x}\, \frac{d\vp}{du} \quad \to \quad
\frac{dy}{dx}= 3\a^2\,\frac{y^2 \bigl(f_1 - 2{\bar f}_2 xy\bigr)}{x
\bigl\lb M_1 - 2{\bar M}_2 x + y (f_1 - 2{\bar f}_2 xy)\bigr\rb} \; ,
\lab{y-x-curve}
\ee
and the solution of Eq.\rf{y-x-curve} is obtained as power series w.r.t. 
a very small parameter $\a^2$:
\br
y(x) = y_0 + \a^2 y_1 (x;y_0) + {\rm O}(\a^4) \phantom{aaaaaaaaaaaaaaaaa}
\lab{y-0} \\
y_1 (x;y_0) = 3 y_0^2 \Bigl\{ \frac{f_1}{M_1 +f_1 y_0} \log x +
\phantom{aaaaaaaaaaaaaaaaa}
\nonu \\
\(\frac{{\bar f}_2 y_0}{{\bar M}_2 + {\bar f}_2 y_0^2} - \frac{f_1}{M_1 +f_1 y_0} \)
\log \Bigl\lb M_1 +f_1 y_0 - 2x ({\bar M}_2 + {\bar f}_2 y_0^2)\Bigr\rb\Bigr\}\; .
\lab{y-1}
\er
Henceforth we will keep $y_0 \equiv e^{-\a\vp_0}$ as zero-order term in the
$\a^2$-expansion of $y(x)$ \rf{y-0}.
%%%%%%%%%%%%%%%%%%%%%%%

Following the geometric description of two-field inflationary trajectories
in Ref.\ct{tegmark}, Eqs.\rf{y-x-curve}-\rf{y-1} mean that the present
inflationary curve $y = y(x)$ % (or $\vp = \vp (u)$
has a very small curvature:
\be
K \equiv \frac{d^2 y}{dx^2}\,\Bigl\lb 1 + \bigl(\frac{dy}{dx}\bigr)^2\Bigr\rb^{-3/2}
= {\rm O}(\a^2)
\lab{}
\ee
% or order ${\rm O}(\a^2)$ 
(\textsl{i.e.}, very small turn rate) because
of the small coupling parameter $\a$, and thus is close to an inflation
along the $u$-field direction.

%%%%%%%%%%%%%%%%%%%%%%%

Taking into account \rf{vp-sol}-\rf{vp-sol-1} (or their counterparts
\rf{y-0}-\rf{y-1}) and (recall
$\bigl(x \equiv e^{-u/\sqrt{3}},\, y \equiv e^{-\a\vp}\bigr)$):
\br
V_{\vp\vp}=\a^2 \bigl\lb - f_1 xy + 4 {\bar f}_2 x^2 y^2\bigr\rb \;\; ,\;\;
V_{\vp u}=\frac{\a}{\sqrt{3}}\bigl\lb - f_1 xy + 4 {\bar f}_2 x^2 y^2\bigr\rb \; ,
\nonu \\
V_{uu}=\frac{1}{3}\bigl\lb -M_1 x + 4{\bar M}_2 x^2 
- f_1 xy + 4 {\bar f}_2 x^2 y^2\bigr\rb \; ,
\lab{V-2nd-der}
\er
the standard slow-roll parameters: 
% ($\eps = - \frac{\Hdot}{H^2}$ and $\eta = \eps - \frac{Hddot}{2H\Hdot}$):
\br
\eps = - \frac{\Hdot}{H^2} = \frac{V_{\vp}^2 + V_u^2}{V^2} = 
\Bigl(1+\bigl(\frac{d\vp}{du}\bigr)^2\Bigr)\,\(\frac{V_u}{V}\) \bgv_{\vp = \vp(u)}
\; ,
\nonu \\
\eta = \eps - \frac{\Hddot}{2H\Hdot} = 
2 \frac{V_{\vp\vp}V_{\vp}^2 + V_{uu}V_u^2 +2 V_{u\vp}V_u V_{\vp}}{
V(V_{\vp}^2 + V_u^2)} \bgv_{\vp = \vp(u)} \; ,
\lab{slow-roll-param}
\er
can be written as functions of $(x,y)$
% $\bigl(x \equiv e^{-u/\sqrt{3}},\, y \equiv e^{-\a\vp}\bigr)$ 
upto terms of order ${\rm O}(\a^2)$:
\br
\eps = \frac{4\xi^2 (x,y) \Bigl(\h - \xi(x,y)\Bigr)^2}{3 
\Bigl\lb \Bigl(\h - \xi(x,y)\Bigr)^2 + \frac{1}{4}\bigl( X(y)-1\bigr)\Bigr\rb^2}
+ {\rm O}(\a^2) \; ,
\nonu \\
|\eta| = \frac{2\xi(x,y) \Bigl(1 - 4\xi(x,y)\Bigr)}{3
\Bigl\lb \Bigl(\h - \xi(x,y)\Bigr)^2 + \frac{1}{4}\bigl( X(y)-1\bigr)\Bigr\rb}
+ {\rm O}(\a^2) \; ,
\lab{slow-roll-param-1}
\er
where we used the short-hand notations:
\be
\xi(x,y) \equiv x\,\frac{{\bar M}_2 + {\bar f}_2 y^2}{M_1 + f_1 y} \quad ,\quad
X(y) \equiv 
\frac{8\L_0 \bigl({\bar M}_2 + {\bar f}_2 y^2\bigr)}{\bigl(M_1 + f_1 y\bigr)^2} \; .
\lab{xi-X-def}
\ee

The end of the inflation occurs for $u=u_{\rm f}$ or 
$x_{\rm f} = e^{- u_{\rm f}/\sqrt{3}}$ when $\eps \equiv \eps(x,y)$ in
\rf{slow-roll-param-1} becomes $\eps (x_{\rm f}, y_0) = 1$ (discarding terms
of order ${\rm O}(\a^2)$):
\be
% \xi_{\rm f} \equiv 
% e^{- u_{\rm f}/\sqrt{3}} \,\frac{{\bar M}_2 + {\bar f}_2 y_0^2}{M_1 + f_1 y_0}
x_f = \frac{M_1 + f_1 y_0}{{\bar M}_2 + {\bar f}_2 y_0^2}
\; \frac{1}{2(1+2/\sqrt{3})}
\llb 1+\frac{1}{\sqrt{3}} - \sqrt{1/\sqrt{3} - (1+2/\sqrt{3}) (X (y_0)-1)}\rrb
+ {\rm O}(\a^2) \; ,
\lab{x-f}
\ee
with $X (y_0)$ as in \rf{xi-X-def} for $y_0 \equiv e^{-\a\vp_0}$.

Using relation \rf{DE-value} for $\L_0$ and ignoring the very small value
of $\L_{\rm DE}$, $X (y_0)$ becomes:
\be
X (y_0)= \frac{\Bigl(1+y_0^2 f_2/M_2\Bigr)}{\bigl(1+y_0 f_1/M_1\bigr)^2}\;
\(1+\frac{M_2 f_1^2}{M_1^2 f_2}\)  \quad , \quad
1 \leq X (y_0) \leq 1+\llb 3(1+2/\sqrt{3})\rrb^{-1} \; ,
\lab{X-eq}
\ee
where the upper limit in the last inequality is implied by the
reality of $x_{\rm f}$ \rf{x-f}. Let us particularly note, that the
expression and inequality \rf{X-eq} explicitly exhibits the difference 
between the present two-field
$(u,\vp)$ inflation (even when discarding the $\a^2$-corrections) and its
single field $u$-inflation counterpart, where 
$X \equiv 8\L_0\,\frac{{\bar M}_2}{M_1^2} \simeq 1$ (upon ignoring 
$\L_{\rm DE}$), cf. Ref.\ct{Benisty:2019tno}.

For the number of e-folds using $\frac{d\vp}{du}= V_{\vp}/V_u$ we have
\be
\cN = \h\int \frac{V\bigl(V_{\vp} d\vp + V_u du\bigr)}{V_{\vp}^2 + V_u^2} =
\h \int_{u_i}^{u_f} du \frac{V}{V_u}\bgv_{\vp = \vp(u)} \; ,
\lab{e-folds-def}
\ee
which can be represented as:
\be
\cN = \frac{3}{2} \int_{x_f}^{x_i} \frac{dx}{x^2} \;
\frac{2\L_0 - x \bigl(M_1+f_1 y(x)\bigr) 
+ x^2 \bigl({\bar M}_2 + {\bar f}_2 y^2(x)\bigr)}{M_1+f_1 y(x)
-2x \bigl({\bar M}_2 + {\bar f}_2 y^2(x)\bigl)} \; ,
\lab{e-fold-eq}
\ee
where $x_i \equiv e^{-u_i/\sqrt{3}}$ and $x_f \equiv e^{-u_f/\sqrt{3}}$
correspond to start and end of inflation.

Using the $\a^2$-expansion for $y(x)$ \rf{y-0} we obtain from \rf{e-fold-eq},
using again relation \rf{DE-value} for $\L_0$ (and discarding the very small
$\L_{\rm DE}$) as well as the short-hand notation for $X(y_0)$ \rf{X-eq}:
\be
\cN \equiv \cN(x_i, y_0) = \frac{3}{8} X (y_0) \Bigl(1/\xi_{\rm i} - 1/\xi_{\rm f}\Bigr)
- \frac{3}{4}\Bigl(2 - X (y_0)\Bigr) \log\frac{\xi_{\rm f}}{\xi_{\rm i}}
+ \frac{3}{8} \Bigl(X (y_0) - 1\Bigr) \log\frac{1-2\xi_{\rm f}}{1-2\xi_{\rm i}} 
+ {\rm O}(\a^2)\; ,
\lab{e-fold-sol}
\ee
where $\xi_{\rm f} \equiv \xi(x_{\rm f}, y_0)$ and
$\xi_{\rm i} \equiv \xi(x_{\rm i}, y_0)$ with $\xi(x,y)$ as in \rf{xi-X-def} and 
$x_f$ given by Eq.\rf{x-f}. 

\begin{figure}[t!]
\centering
\includegraphics[width=0.55\textwidth]{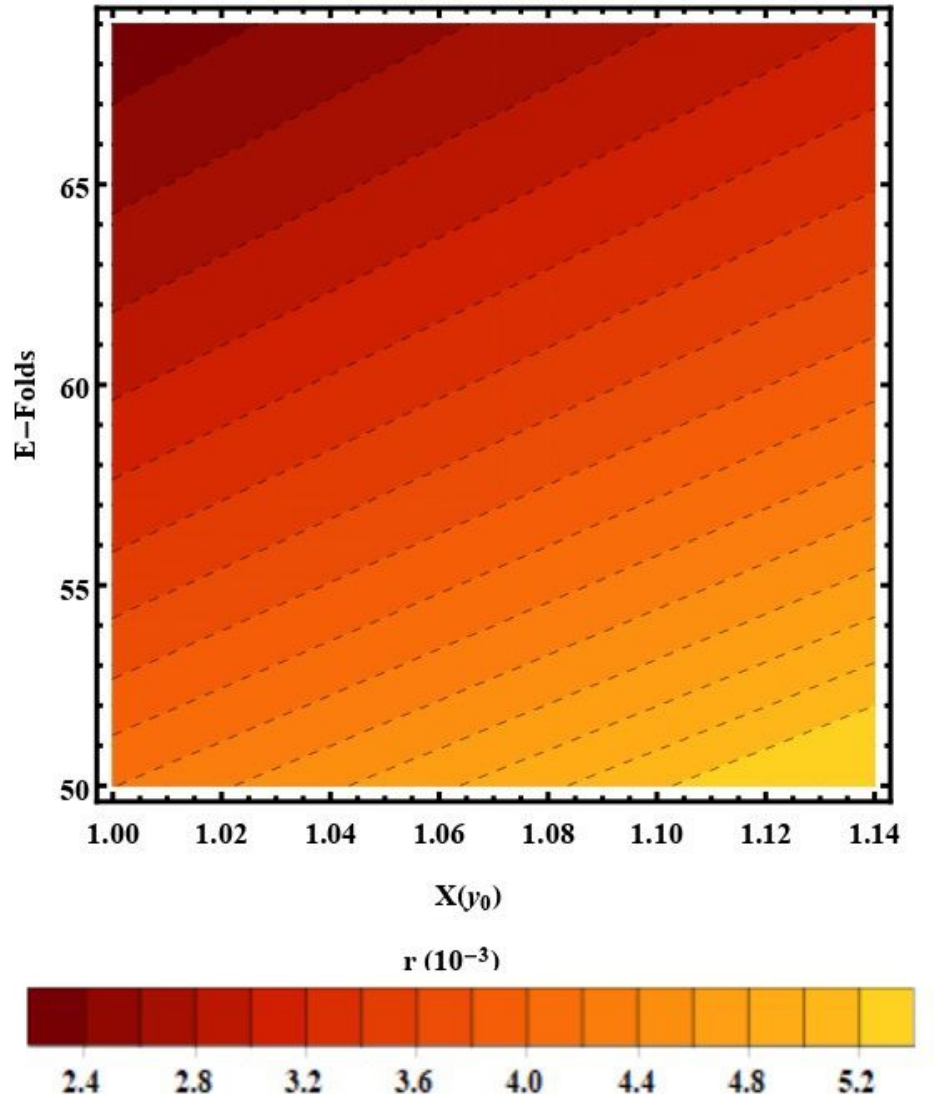}
\includegraphics[width=0.6\textwidth]{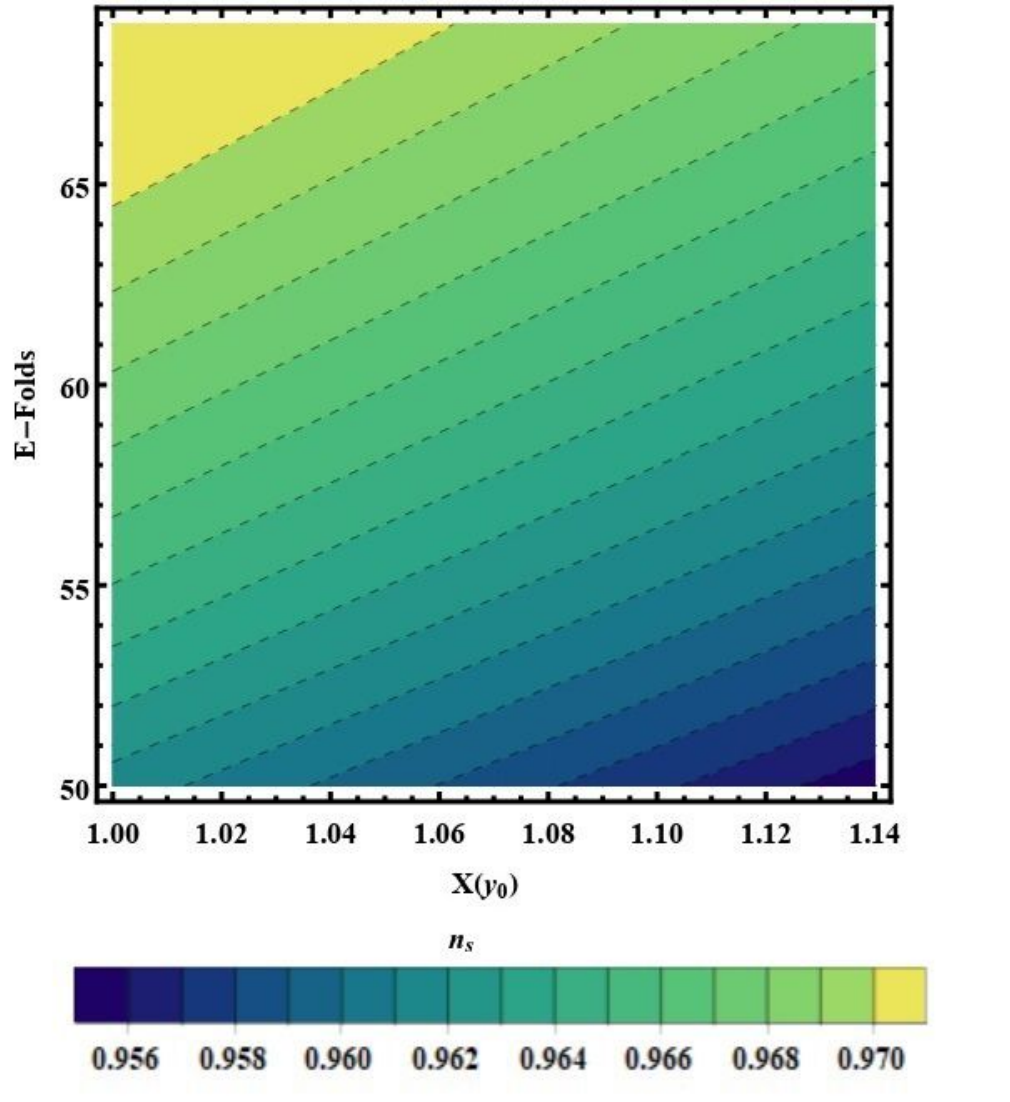}
\caption{The predicted values of the $r$ and $n_s$ for different $e$-foldings, with different values of $X (y_0)$ as defined in Eq. \rf{xi-X-def}. In the upper panel the contour presents the values of $r$ while in the lower panel the contour presents the values of $n_s$. The different values of the $r$ and $n_s $ are compatible with the observational data.}
\label{fig2}
\end{figure}

Through the expressions for the slow-roll parameters \rf{slow-roll-param-1}
and the number of e-folds \rf{e-fold-sol}
one can calculate the values of the scalar spectral index and 
the tensor-to-scalar ratio, respectively, as functions of the number of e-folds  
\cite{Nojiri:2019kkp,Dalianis:2018frf}:
\be
r \approx 16 \eps(x_i,y_0)\quad ,
\quad n_s \approx 1- 6\eps(x_i,y_0) + 2\eta(x_i,y_0) \quad , \quad 
{\rm with} \;\; \cN = \cN(x_i,y_0) \; .
\lab{r-ns-eqs}
\ee
Unlike the case of single-field $u$-inflation \ct{Benisty:2019tno}, 
now the results will
parametrically depend on the value of $y_0 = e^{-\a\vp_0}$ through $X(y_0)\neq 1$. 
Fig.\ref{fig2} contains  contour plots for the predicted values 
of the $r$ and $n_s$ for different $e$-foldings, with different allowed 
values of $X (y_0)$.
All of the observables fit within the PLANCK data with the constraints
\cite{Akrami:2018odb}:
\be
0.95 < n_s < 0.97, \quad r < 0.064 \; ,
\lab{planck-constraints}
\ee
however in future the observables are going to be more constrained. 
The future Euclid and SPHEREx missions or the
BICEP3 experiment are expected to provide experimental evidence to test those 
predictions.

%%%%%%%%%%%%%%%%%%%%%%%
%%%%%%%%%%%%%%%%%%%%%

%%%%%%%%%%%%%%%%%%%%%%%%%%%%%%%%%%%%%%%%%%%%%%%%%%%%%%%%%%%%%%%%%%%%%%%%%%%%%%%
%%%%%%%%%%%%%%%%%%%%%%%%%%%%%%%%%%%%%%%%%%%%%%%%%%%%%%%%%%%%%%%%%%%%%%%%%%%%%%%
\section{Conclusions}

In the present paper we propose a simple model of modified gravity 
interacting with a single scalar field $\vp$ weakly coupled via exponential potential. 
The construction is based on the formalism of non-Riemannian spacetime
volume-elements. In addition, the structure of the initial action is specified
by the requirement of invariance under global Weyl-scale symmetry. 
Since we are employing the second order (metric) formalism for the gravity
part of the action, the transition from the original frame with the non-Riemannian 
volume elements to the physical Einstein frame creates dynamically a 
second canonical scalar field $u$ accompanied by the emergence of several free
integration constants. All this leads to the dynamical generation of a
non-trivial two-scalar-field potential $U_{\rm eff}(u,\vp)$.
We show that the dynamically created 
field $u$ serves as an inflaton field in the region of slow-roll inflation
which is a flat region of the potential $U_{\rm eff}(u,\vp)$ for large $u$. 
Furthermore, we explicitly show that $U_{\rm eff}(u,\vp)$ possesses a stable
very low lying minimum as function of $(u,\vp)$ appropriate to describe the
dark energy dominated ``late'' universe with a very small dark energy density.

We show that the pertinent slow-roll inflationary curve in $(u,\vp)$-space
$\vp = \vp(u)$ has very small curvature in a sense that $\vp$ changes very
little during the inflationary evolution of $u$. We find an analytic solutions 
to the observables -- scalar to tensor ratio $r$ and the scalar spectral
index $n_s$, which fit the known observational data. 
Beyond this prediction, the interaction between the two scalarfields leads to 
different values of the observables, 
which in future experiments will be constrained more efficiently. In what
follows we intend to study the two-field inflationary evolution curve $\vp = \vp(u)$
relaxing the assumption of weakly coupled $\vp$, especially in the context of
the non-Gaussianity feature in CMB (for recent review, see \ct{CMB} and references
therein) , which might provide more justification for 
the viability of the present model. 
%%%%%%%%%%%%%%%%%%

\section{Acknowledgments}
We gratefully acknowledge support of our collaboration through the Exchange 
Agreement between Ben-Gurion University, Beer-Sheva, Israel and Bulgarian 
Academy of Sciences, Sofia, Bulgaria. E.N. and S.P. are thankful for support 
by Contract DN 18/1 from Bulgarian National Science Fund. D.B., E.G. and E.N. 
are also partially supported by COST Actions CA15117, CA16104 and CA18108. 
D.B., E.N. and S.P. acknowledge illuminating discussions with Lilia Anguelova.
%%%%%%%%%%  CHANGE-BEGIN  %%%%%%%%%%%%
Finally, we thank the referee for constructive remarks contributing to
improvement of the presentation.
%%%%%%%%%%  CHANGE-END  %%%%%%%%%%%%

\end{document}